OPEN

# Data Descriptor: Climatologies at high resolution for the earth's land surface areas

Dirk Nikolaus Karger[1,2], Olaf Conrad[3], Jürgen Böhner[3], Tobias Kawohl[3], Holger Kreft[4], Rodrigo Wilber Soria-Auza[4,5], Niklaus E. Zimmermann[2], H. Peter Linder[1] & Michael Kessler[1]

High-resolution information on climatic conditions is essential to many applications in environmental and ecological sciences. Here we present the CHELSA (Climatologies at high resolution for the earth's land surface areas) data of downscaled model output temperature and precipitation estimates of the ERA-Interim climatic reanalysis to a high resolution of 30 arc sec. The temperature algorithm is based on statistical downscaling of atmospheric temperatures. The precipitation algorithm incorporates orographic predictors including wind fields, valley exposition, and boundary layer height, with a subsequent bias correction. The resulting data consist of a monthly temperature and precipitation climatology for the years 1979–2013. We compare the data derived from the CHELSA algorithm with other standard gridded products and station data from the Global Historical Climate Network. We compare the performance of the new climatologies in species distribution modelling and show that we can increase the accuracy of species range predictions. We further show that CHELSA climatological data has a similar accuracy as other products for temperature, but that its predictions of precipitation patterns are better.

| | |
|---|---|
| **Design Type(s)** | data integration objective • modeling and simulation objective |
| **Measurement Type(s)** | temperature of air • hydrological precipitation process |
| **Technology Type(s)** | data acquisition system |
| **Factor Type(s)** | |
| **Sample Characteristic(s)** | Earth • planetary atmosphere |

[1]Department of Systematic and Evolutionary Botany, University of Zurich, Zollikerstrasse 107, Zurich 8008, Switzerland. [2]Swiss Federal Research Institute WSL, Zürcherstr 111, Birmensdorf 8903, Switzerland. [3]Institute of Geography, University of Hamburg, Bundesstrasse 55, Hamburg 20146, Germany. [4]Biodiversity, Macroecology & Conservation Biogeography Group, University of Göttingen, Göttingen 37077, Germany. [5]Asociación Armonía, Av. Lomas de Arena # 400, Zona Palmasola, Santa Cruz de la Sierra 10260, Bolivia. Correspondence and requests for materials should be addressed to D.N.K. (email: dirk.karger@wsl.ch).

## Background & Summary

High-resolution climate data are essential to many applications in environmental and ecological sciences. Whereas many studies in these fields are conducted at a resolution of ~1 km$^2$, state-of-the-art global climate reanalyses often only represent climatic variation at spatial resolutions of 0.25°–1° (ca. 25–100 km at the equator). The gap between these spatial scales may be bridged using satellite data (CHIRPS[1], TRMM[2,3]) and statistical downscaling[4–7] for a specific region of interest and/or interpolation methods applied to meteorological station data (WorldClim[8], CRU[9], GPCC[10], PRISM[11]). Climatologies based on satellites or statistical downscaling, considered superior to interpolated data[12] for ecological applications, are currently either not available on a global scale or are still too coarse to reflect the small scale patterns needed in ecological studies. While interpolated datasets[8] often perform well in matching precipitation or temperature of the stations from which they are produced, they often fail to accurately predict patterns between stations. This is particularly problematic in highly variable terrain with low station density[13]. Whereas some interpolated datasets use elevation as a predictor (e.g., WorldClim[8]) and observations such as the Global Historical Climate Network (GHCN)[14,15] to achieve a high-resolution prediction, it is also possible to use predictors from global circulation models (e.g., from the National Center for Atmospheric Research (NCEP)[16], or the European Centre for Medium-Range Weather Forecast (ECMWF) climatic reanalysis interim (ERA-Interim)[17]).

Although interpolation and statistical downscaling approaches may also integrate land-surface predictors such as elevation, slope or aspect, satisfactory results still require a more or less regular distribution of meteorological stations and a proper representation of topo-climatic settings[13]. However, the global distribution of meteorological stations is highly biased by funding and accessibility, leading to a poor representation of climatic variability in mountainous regions or areas with intact lowland rainforest, such as the Amazon or Congo basin[8,13]. On a global scale, it is also difficult to find generally valid transfer functions between predictors and the climatic variable of interests, especially for highly non-linear phenomena such as precipitation. Statistical downscaling is problematic[18], especially on a global scale, due to temporal variation in the spatial distribution of weather stations. Although measurements for a given predictor might be available in a given month, they might be absent in another, leading to a generally high heterogeneity of the underlying climate records when time series of precipitation need to be calculated. While this does not affect static predictors such as elevation, slope, or aspect, statistical downscaling becomes especially problematic when highly dynamic predictors such as wind fields need to be integrated. The heterogeneity in the temporal and spatial distribution of such dynamic factors can also lead to spurious correlations in specific months or in specific regions, which can severely influence regression model parameters. When specific predictors, such as windward or leeward mountain sides[19,20] change over the course of the year, the location of the climatic records does not change accordingly. Therefore, regression-based downscaling might, for example, detect a significant negative relationship between a station on the windward site of a mountain for one month, and a positive relationship for another, although atmospheric physics would always predict a positive relationship. Due to this problem, statistical downscaling and interpolation methods have often been applied to single regions[11], while a global model is lacking.

For ecological applications, the representation of the temporal and spatial variability of temperature and precipitation is, however, extremely important to infer ecological niches, growing seasons, species migrations, or small scale species distribution. Errors in the underlying climatic dataset at this small spatial scale can easily inflate in such studies[13], which calls for an improvement of climatic information available for such analyses.

To overcome the problem of heterogeneous spatial and temporal distribution of meteorological station data, we use a Model Output Statistics algorithm for data provided from the ERA-Interim reanalysis[17] which we correct using gauge-derived products from the GPCC[10] and the GHCN[14,15] datasets. The results are improved climatologies for precipitation and temperature at high spatial resolution for environmental and ecological studies, which might prove valuable in varied scientific applications that rely on a good representation of small scale precipitation and temperature patterns.

## Methods

### Calculation of monthly temperature and precipitation values

ERA-Interim (developed at the European Centre for Medium-Range Weather Forecast, ECMWF), simulates six-hourly large-scale atmospheric fields for 60 pressure levels between 1,000 and 1 hPa globally with a horizontal resolution of 0.75° lat/long (T255)[17,21,22]. Since the ERA-Interim reanalysis combines modelling results with ground and radiosonde observations as well as remote sensing data using a data assimilation system, the free-atmospheric and surface fields can be considered as the best approximation of the current large-scale atmospheric situation for every time step. Several studies show that ERA-Interim adequately captures the variability of relevant free-air meteorological parameters, even over complex terrain[23–25].

### Temperature

Spatial variation in temperature is to a large degree determined by the vertical state of the troposphere and thus, if not affected by inversion layers, temperature decreases with increasing altitude[26,27]. The long term mean hypsometric temperature gradient covered in the ERA-Interim data accurately reflects the vertical distribution of moist- or dry-adiabatic lapse rates[20]. Typical temperature lapse rates are in the order of −0.4 to −0.8 K/100 m with a characteristic seasonality. The corresponding temperature distribution pattern in the free atmosphere[28] can be assumed to be directly related to surface elevation[19].

For our downscaling approach of mean monthly temperatures, we used the monthly means of daily mean temperature derived from six-hourly synoptic data from ERA-Interim. Temperature lapse rates were calculated from the ERA-Interim for pressure levels from 1,000 to 300 hPa, using linear regression for each ERA-Interim grid cell. We then interpolated temperature to sea level using the derived lapse rates. Sea level temperatures were then interpolated between grid cells using B-spline interpolation, and then projected back on the elevational surface of the digital elevation model using the equation:

$$t = \Gamma_d * elev + t_0 \quad (1)$$

where $t$ equals the temperature at a given elevation, $\Gamma_d$ equals the lapse rate, *elev* equals elevation at 30 arc sec. from the Global Multi-resolution Terrain Elevation Data 2010 (GMTED2010)[29] of the United States Geological Survey (USGS) and the National Geospatial-Intelligence Agency (NGA), and $t_0$ equals the interpolated temperature at sea level.

## Maximum and minimum temperatures

Maximum ($t_{max}$) and minimum ($t_{min}$) temperatures were calculated using climatological aided interpolation[30]. For that we used the mean monthly temperature values ($t$) and added, or subtracted the maximum or minimum daily temperature derived from the three-hourly data of minimum or maximum temperature since previous post processing data fields in ERA-Interim:

$$t_{max} = t + \Delta t_{era_max} \quad (2)$$

$$t_{min} = t - \Delta t_{era_min} \quad (3)$$

where $\Delta t_{era_max}$ and $\Delta\ t_{era_min}$ are the respective differences between maximum and minimum temperatures interpolated to 30 arc sec resolution using B-spline interpolation from the mean monthly temperatures ($t$).

## Precipitation

Elevation is one of the main topo-climatic drivers of vertical precipitation gradients, but the relation between elevation and precipitation can be idiosyncratic[19,31–36]. In the convective regimes of the tropics, precipitation amounts commonly increase up to the condensation level at about 1,000–1,500 m above the ground surface, and the exponentially decreasing air moisture content in the mid- to upper troposphere results in a corresponding drying above the condensation level of tropical convection cluster systems (non-linear precipitation lapse rates)[37]. Likewise, negative lapse rates typically occur in the extremely dry polar climates. At mid-latitudes and in the subtropics, the frequent or even prevalent advection of moisture bearing air to high altitudes leads to increasing precipitation with increasing elevation. Consequently, the summits of high mountain ranges such as the Alps[38] may have high rainfall, and this leads to linear precipitation lapse rates[39]. The reduced precipitation at lower elevations is due, firstly, to the evaporation of rain drops when falling through non-saturated, lower-air levels. Secondly, the vertical precipitation gradient in high mountain ranges is often increased due to the diurnal formation of autochthonous upslope breezes. This upward flow of air intensifies cloud and precipitation formation in upper slope positions whilst the subsiding branch of these autochthonous local circulation systems along the valley axis leads to cloud dissolution and a corresponding reduction of precipitation rates in the valley bottoms. We approximated such orographic precipitation effects and used them as a parameter for the CHELSA precipitation downscaling algorithm (Fig. 1) as explained below.

## Wind effect correction

Orographic precipitation patterns[40] caused by the uplift of moist air currents at the windward side of a mountain range and the intimately related rain shadow effect on leeward sides induced by the blockage of moisture-bearing air are most common effects influencing small-scale precipitation patterns[38,40–43]. Based on the assumption that the windward impact on the precipitation intensity depends on the prevailing wind direction at any given elevation of an orographic barrier, we used a wind index[19,20] to account for the expected higher precipitation at the windward sites of an orographic barrier.

We used u-wind and v-wind components at the 10-m level of ERA-Interim as underlying wind components. These two wind components were interpolated to the CHELSA grid resolution using a B-spline interpolation. As the calculation of a windward leeward index (hereafter: wind effect) requires a projected coordinate system, both wind components were projected to a world Mercator projection and then combined to a directional grid. The wind effect $H$ with windward component $H_W$ and the leeward

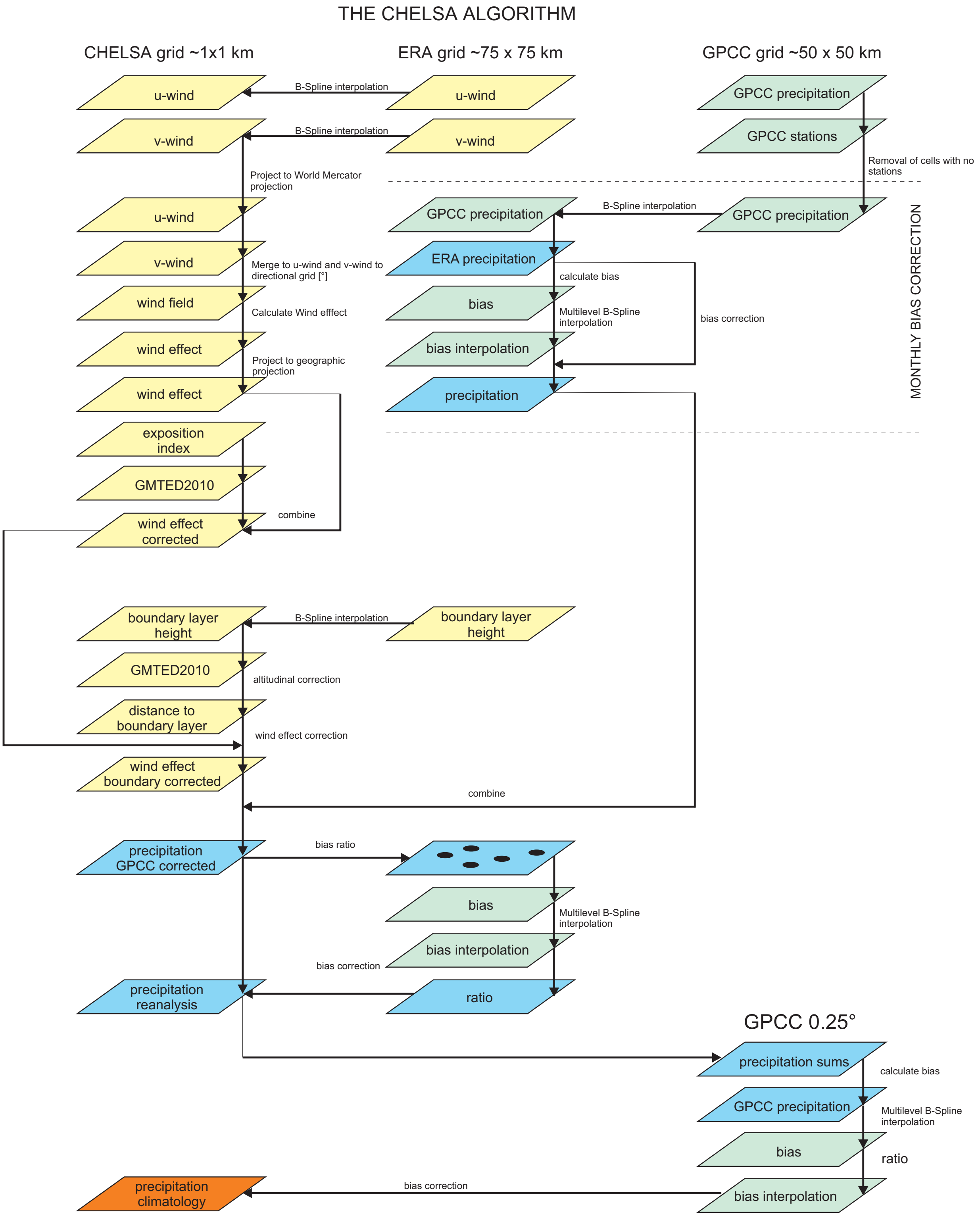

**Figure 1. Workflow of the CHELSA model output statistics and interpolation algorithm for precipitation data.** Resulting raster datasets (parallelograms) from each calculation step (arrows) are shown for each step of the algorithm. Predictor variables are indicated in yellow, raster datasets of the dependent variable (precipitation) are indicated in blue, and bias correction raster datasets are indicated in green. The final climatological product is indicated in orange.

component $H_L$ were then calculated using:

$$H_{\mathrm{W}} = \frac{\sum_{i=1}^{n} \frac{1}{d_{\mathrm{WH}i}} tan^{-1}\left(\frac{d_{\mathrm{WZ}i}}{d_{\mathrm{WH}i}}\right)}{\sum_{i=1}^{n} \frac{1}{d_{\mathrm{LH}i}}} + \frac{\sum_{i=1}^{n} \frac{1}{d_{\mathrm{LH}i}} tan^{-1}\left(\frac{d_{\mathrm{LZ}i}}{d_{\mathrm{LH}i}}\right)}{\sum_{i=1}^{n} \frac{1}{d_{\mathrm{LH}i}}} \tag{4}$$

$$H_{\mathrm{L}} = \frac{\sum_{i=1}^{n} \frac{1}{d_{\mathrm{WH}i}} tan^{-1}\left(\frac{d_{\mathrm{LZ}i}}{d_{\mathrm{WH}i}}\right)}{\sum_{i=1}^{n} \frac{1}{d_{\mathrm{LH}i}}} \tag{5}$$

where $d_{\mathrm{WH}i}$ and $d_{\mathrm{LH}i}$ refer to the horizontal distances in windward and leeward direction and $d_{\mathrm{WZ}i}$ and $d_{\mathrm{LZ}i}$ are the corresponding vertical distances compared with the considered raster cell. The second summand in equation (4) accounts for the leeward impact of previously traversed mountain chains. The horizontal distances in equation (5) lead to a longer-distance impact of leeward rain shadow. The final wind-effect parameter, which is assumed to be related to the interaction of the large-scale wind field and the local-scale precipitation characteristics, is calculated as $H = H_{\mathrm{L}} \times H_{\mathrm{W}}$ and generally takes values between 0.7 for leeward and 1.3 for windward positions[19]. Equation (3) and equation (4) were applied to each grid cell at the CHELSA resolution in a world mercator projection.

### Valley exposition correction

Although the wind effect algorithm can distinguish between the windward and leeward sites of an orographic barrier, it cannot distinguish extremely isolated valleys in high mountain areas. Such dry valleys are situated in areas where the wet air masses flow over an orographic barrier and are prevented from flowing into deep valleys. To account for these effects, we used a variant of equation (4) and equation (5) with a linear search distance of 300 km in steps of 5° from 0° to 355° circular for each grid cell. The calculated leeward index was then scaled towards higher elevations using:

$$E = H_L^{\frac{elev}{c}} \tag{6}$$

which rescales the strength of the exposition index relative to elevation (*elev*) from GMTED2010, and gives valleys at high elevations larger wind isolations ($E$) than valleys located at low elevations. The correction constant $c$ was set to 9,000 m to include all possible elevations of the DEM. The constant has been set to 9,000 m as values of $elev > c$ could lead to a reverse relationship between *elev* and $H_L$. Additionally, a prior sensitivity analysis indicated that downscaled precipitation with $c = 9{,}000$ m has a better fit with precipitation measured at the stations (GHCN stations) than values of $c > 9{,}000$ m. We therefore choose to set $c$ conservatively to 9,000 m.

### Boundary layer correction

Orographic precipitation effects are less pronounced just above the surface, as well as in the free atmosphere above the planetary boundary layer[11,44,45]. The highest impact of orography is considered just at the boundary layer height where the airflow interacts with the terrain. While former studies used single ERA pressure levels, known to represent the main wind field patterns in a specific area[20], the pressure level representing the prevailing wind directions at the boundary layer is usually not known a priory on a global basis. We therefore used the boundary layer height $B$ from ERA-Interim as indicator of the pressure level that has the highest contribution to the wind effect. The boundary layer height has been interpolated to the CHELSA resolution using a B-spline interpolation. The wind effect grid $H$ containing the windward ($H_{\mathrm{W}}$) and leeward ($H_{\mathrm{L}}$) index values was then proportionally distributed to all grid cells falling within a respective 0.75° grid cell using:

$$H_{WB} = \frac{H_W}{1 - \left(\frac{|d| - d_{max}}{c}\right)} \tag{7}$$

$$H_{LB} = \frac{H_L}{1 - \left(\frac{|d| - d_{max}}{c}\right)} \tag{8}$$

with:

$$d = elev - B \tag{9}$$

With $d$ being the distance between a grid cell and the boundary layer height $B$, $d_{max}$ being the maximum distance between the boundary layer height $B$ and all grid cells at the CHELSA resolution falling within a respective 0.75° grid cell, $c$ being a constant of 9,000 m, and *elev* being the respective elevation from GMTED2010.
with:

$$B = B_{ERA} + elev_{ERA} + f \tag{10}$$

$B$ being the height of the monthly means of daily mean boundary layer from ERA-Interim, $elev_{ERA}$ being the elevation of the ERA-Interim grid cell, and $f$ being a constant of 500 m which takes into account that the level of highest precipitation is not necessarily at the lower bound of the boundary layer, but slightly higher[44,45]. Similar to the $c$ value in equations (6–8) we used a prior sensitivity analysis that varied $f$ in steps of 50 m to determine the impact of $f$ on the modelled precipitation values. Values of 500 $>f<$ 500 showed to a lower fit between modelled precipitation and precipitation measured at stations.

### Precipitation data from ERA-Interim

For accumulated parameters (total monthly precipitation), we used the monthly means of daily forecast accumulations of total precipitation initialized at the synoptic hours 0:00 and 12:00. To calculate monthly precipitation sums, we added the synoptic monthly means at time 0:00, step 12 and time 12:00, step 12 and multiplied it by the number of days in the respective month.

### Bias correction of ERA-Interim data using GPCC and GHCN data

Model-generated estimates of the surface precipitation are extracted from short range forecasts, which vary with forecast length. This drift in the short-range forecasts can be a problem for monthly and climatic means[46]. One very common approach is to calculate the difference between baseline precipitation from the GCM and the observed precipitation and apply this 'factor of change' to historically observed time series to generate a synthetic time series[47–49]. We therefore performed three steps of bias correction.

### Monthly bias correction

We applied the monthly bias correction before the downscaling of the precipitation data on the ERA-interim precipitation values $p_{\mathrm{ERA}}$ directly[49]. To this end, we used the monthly values $p_{\mathrm{GPCC}}$ of the gridded GPCC dataset[10] to calculate the monthly bias $R_m$ caused by the ERA-Interim parametrization, and the excessive or insufficient precipitation of the forecast algorithm[46] for each month from Jan. 1979–Dec. 2013 using:

$$R_m = \frac{p_{GPCC}}{p_{ERA}} \tag{11}$$

We only used grid cells with meterological stations present for $R_m$. The forecast algorithm used to produce the precipitation amounts for ERA-Interim exhibits a considerable spin up—spin down effect (too much or too less precipitation), that has a coherent spatial structure, with a larger bias over high elevation terrain, or specific land forms such as tropical rainforests[46]. Based on this observation, we assumed that grid cells without stations share a similar bias as their neighbouring stations. To achieve a gap-free grid surface, we interpolated the gaps in the $R_m$ grid using a multilevel B-spline interpolation with 14 error levels to a 0.75° resolution. The gap-free bias correction surface $R_m$ was then multiplied with the ERA-Interim precipitation $p_{ERA}$ to get the bias corrected monthly precipitation sums $p_m$ at 0.75° resolution:

$$p_m = p_{ERA} * R_m \tag{12}$$

### Monthly precipitation including orographic effects

To achieve the distribution of monthly precipitation sums $p$ including orographic effects, we used a linear relationship between the monthly bias corrected precipitation grids at the ERA resolution $p_m$ and the boundary layer corrected wind effect surface H:

$$p = \frac{H}{\overline{H}} * p_m \tag{13}$$

where $\overline{H}$ is the mean wind effect at ERA resolution. By using a linear relationship we archive that the data are to scale, e.g., the precipitation at 0.75° resolution exactly matches the mean precipitation at all 30°sec cells within the range of a 0.75° cell.

### Station bias correction

We used precipitation from a set of meteorological stations from GHCN, MeteoSwiss, and DWD to correct the remaining error between $p$ and $p_{\mathrm{Station}}$. We calculated the bias ratio between $p$ and $p_{Station}$ and interpolated the bias ratio using a multilevel B-spline interpolation with 14 error levels to a 0.1° grid which matches the spatial accuracy of many GHCN stations. The resulting bias surface was then multiplied with $p$ to achieve the final monthly precipitation estimates.

### Climatologies

We calculated the climatologies as the mean monthly sum of precipitation in the years 1979–2013 for each month. As slight errors in the precipitation sums can, however, accumulate over time, we applied an additional bias correction step using the GPCC Climatology Version 2015[50]. We used the cells at which stations are present, calculated the bias between the annual accumulations and the GPCC climatology, and used a multilevel B-spline interpolation of the biases to a 0.25 grid to create the bias surface. This bias surface was then multiplied with the mean annual precipitation sums to create the final climatologies.

## Bioclimatic parameters

From the monthly temperature and precipitation values, we additionally calculated a set of derived parameters often used in ecological applications. These bioclimatic variables are derived variables from the monthly mean, min, max, mean temperature, and mean precipitation values. These variables are specifically developed for species distribution modelling and related ecological applications. They represent annual averages (e.g., mean annual temperature, annual precipitation), seasonality (e.g., annual range in temperature and precipitation), and extreme or limiting environmental factors (e.g., temperature of the coldest and warmest month, and precipitation of the wet and dry quarters). A quarter is defined as the period of three months (1/4 of the year). The procedure strictly followed that of WorldClim[8] and ANUCLIM[51]. The equations used to calculate the bioclimatic variables (where applicable) are:

$t$ = monthly temperature [°C]
$p$ = monthly precipitation [mm]

$$\text{bio1}:\quad \left(\sum_{i=1}^{12} t_i\right)/12 \tag{14}$$

$$\text{bio2}:\quad \left(\sum_{i=1}^{12}(t_{max}-t_{min})\right)/12 \tag{15}$$

$$\text{bio3}:\quad (t*(t_{max}-t_{min})/t_{max}-t_{min})*100 \tag{16}$$

$$\text{bio4}:\quad \left(\sqrt{\frac{1}{11}\sum_{i=1}^{12}\left(t_i-\left(\sum_{i=1}^{12}t_i/12\right)\right)^2}\right)*100 \tag{17}$$

$$\text{bio7}:\quad t_{max}-t_{min} \tag{18}$$

$$\text{bio12}:\quad \sum_{i=1}^{12} t_i \tag{19}$$

$$\text{bio15}:\quad \left(\sqrt{\frac{1}{11}\sum_{i=1}^{12}(p_i-\left(\sum_{i=1}^{12}p_i/12\right))^2}\right)/\left(\sum_{i=1}^{12}p_i\right)/12) \tag{20}$$

Not listed here are the variables which are based on quarters (3 consecutive months) or specific (wettest, driest, warmest, coldest) months.

## Code availability

The codes used to calculate CHELSA climatologies are written in C++ and are included in SAGA Version 2.2.7, freely available at www.saga-gis.org under the GNU public license including the necessary source codes. Calculations were done in SAGA Version 2.2.7 on the 'Science Cloud' cloud computing facility of the University of Zurich www.s3it.uzh.ch/infrastructure/sciencecloud/.

# Data Records

The CHELSA data contains records for monthly mean temperature in °C and precipitation values in mm/month, and derived bioclimatic variables for the reference period 1979–2013 in form of GeoTIFF files. The climatologies available for download have been derived from monthly values of precipitation and temperature. The files are freely available at www.chelsa-climate.org as well as Dryad (Data Citation 1).

The file format is GeoTIFF.

Naming convention:

CHELSA_ <variable> <z-scale>_ <month>_ <Version>_land.tif

variable:
prec = precipitation [mm/month]
temp = monthly mean of daily mean temperature [°C*10]
tmax = monthly mean of daily maximum temperature [°C*10]
tmin = monthly mean of daily minimum temperature [°C*10]

CHELSA_bio <z-scale>_ <bioclim-variable>_ <Version>_land.tif

bioclim-variable:
1 = Annual Mean Temperature [°C*10]
2 = Mean Diurnal Range [°C]
3 = Isothermality
4 = Temperature Seasonality [standard deviation]
5 = Max Temperature of Warmest Month [°C*10]
6 = Min Temperature of Coldest Month [°C*10]
7 = Temperature Annual Range [°C*10]
8 = Mean Temperature of Wettest Quarter [°C*10]

9 = Mean Temperature of Driest Quarter [°C*10]
10 = Mean Temperature of Warmest Quarter [°C*10]
11 = Mean Temperature of Coldest Quarter [°C*10]
12 = Annual Precipitation [mm/year]
13 = Precipitation of Wettest Month [mm/month]
14 = Precipitation of Driest Month [mm/month]
15 = Precipitation Seasonality [coefficient of variation]
16 = Precipitation of Wettest Quarter [mm/quarter]
17 = Precipitation of Driest Quarter [mm/quarter]
18 = Precipitation of Warmest Quarter [mm/quarter]
19 = Precipitation of Coldest Quarter [mm/quarter]

## Technical Validation

To validate the results of the CHELSA algorithm and the different bias correction steps applied, we use a statistical cross-validation, compared the results with several comparable products that are available at comparable spatial and temporal resolution, and independent meteorological station data. The climatologies are validated in two steps. First, as the effects of orographic winds are a non-stationary phenomenon, we show a validation of the time series (hereafter: reanalysis) from which the climatologies are created. Second, we show a validation of the final climatological products which are available for download.

### Cross-validation of the bias correction method using monthly stations

To validate the results of the bias correction method using meteorological station data, we employed a cross-validation approach based on repeated split-resampling. We randomly omitted 20% of the stations for validation and used the remaining 80% for the bias correction by multilevel B-spline interpolation. We repeated the randomization 20 times per month for each respective year from 1979–2013. The mean $R^2$ values after cross-validation ranged between 0.53 and 0.90 with a mean of $R^2 = 0.77$ and a root mean squared error (RMSE) ranging from 30.06–189.12 mm (mean = 54.69 mm) globally throughout the years. A small increase in variance throughout the years can be observed, which might be due to the decrease in the number of stations throughout the years from 13,680 (Jan. 1979) to 1951 (Dec. 2013).

### Validation of the orographic precipitation patterns

The CHELSA algorithm distributes the mean precipitation measured in a grid cell onto the expected precipitation pattern which in turn is calculated based on the wind effect and valley exposition indices. To test whether the inclusion of wind effect and valley exposition patterns produce a higher accuracy, we compared the fit between station data and precipitation in cells of 0.75° spatial resolution (correction step 1) and compared it with the fit between station data and corrected precipitation at 30 arc sec. We performed this comparison in five topographically complex regions (Table 1).

The inclusion of the small-scale orographic effects generally leads to a better fit to the station data in all complex areas in CHELSA compared to WorldClim[8]. The improvement can range from 19.67 % variance explained during the Himalayan wet season (Table 1) to even a decrease in variance of −3.67 % variance explained during August in the Alps. In the majority of cases, however, the variance explained between measured station data and orographically corrected precipitation generally increases during this step of the algorithm. The remaining variance between stations and orographic predictors is finally removed using the subsequent bias correction steps.

| Month | Alps | Andes | Himalaya | Rocky Mnts. | South Africa | Global |
|---|---|---|---|---|---|---|
| 1 | 0.03 | 0.03 | 0.02 | 0.00 | 0.02 | 0.01 |
| 2 | 0.05 | 0.02 | 0.03 | 0.01 | 0.02 | 0.01 |
| 3 | 0.05 | 0.02 | 0.06 | 0.02 | 0.01 | 0.01 |
| 4 | 0.08 | 0.00 | 0.14 | 0.02 | 0.03 | 0.02 |
| 5 | 0.03 | 0.01 | 0.10 | 0.00 | 0.00 | 0.01 |
| 6 | 0.01 | 0.01 | 0.17 | 0.01 | −0.03 | 0.03 |
| 7 | −0.03 | 0.02 | 0.20 | 0.01 | −0.02 | 0.04 |
| 8 | −0.04 | 0.01 | 0.15 | 0.00 | −0.02 | 0.02 |
| 9 | 0.02 | −0.01 | 0.13 | 0.00 | 0.02 | 0.01 |
| 10 | 0.03 | 0.00 | 0.07 | 0.00 | 0.02 | 0.01 |
| 11 | 0.02 | 0.03 | −0.01 | 0.00 | 0.01 | 0.01 |
| 12 | 0.02 | 0.04 | 0.00 | 0.00 | 0.00 | 0.01 |

**Table 1. Difference in $R^2$ values before and after the downscaling from 0.75° resolution to 30 arc sec resolution by means of orographic wind effect correction globally and for five topographic complex regions.** Positive values indicate an increase in fit between meteorological stations and model predictions, negative values indicate an increase in noise. Rocky Mnts, Rocky Mountains.

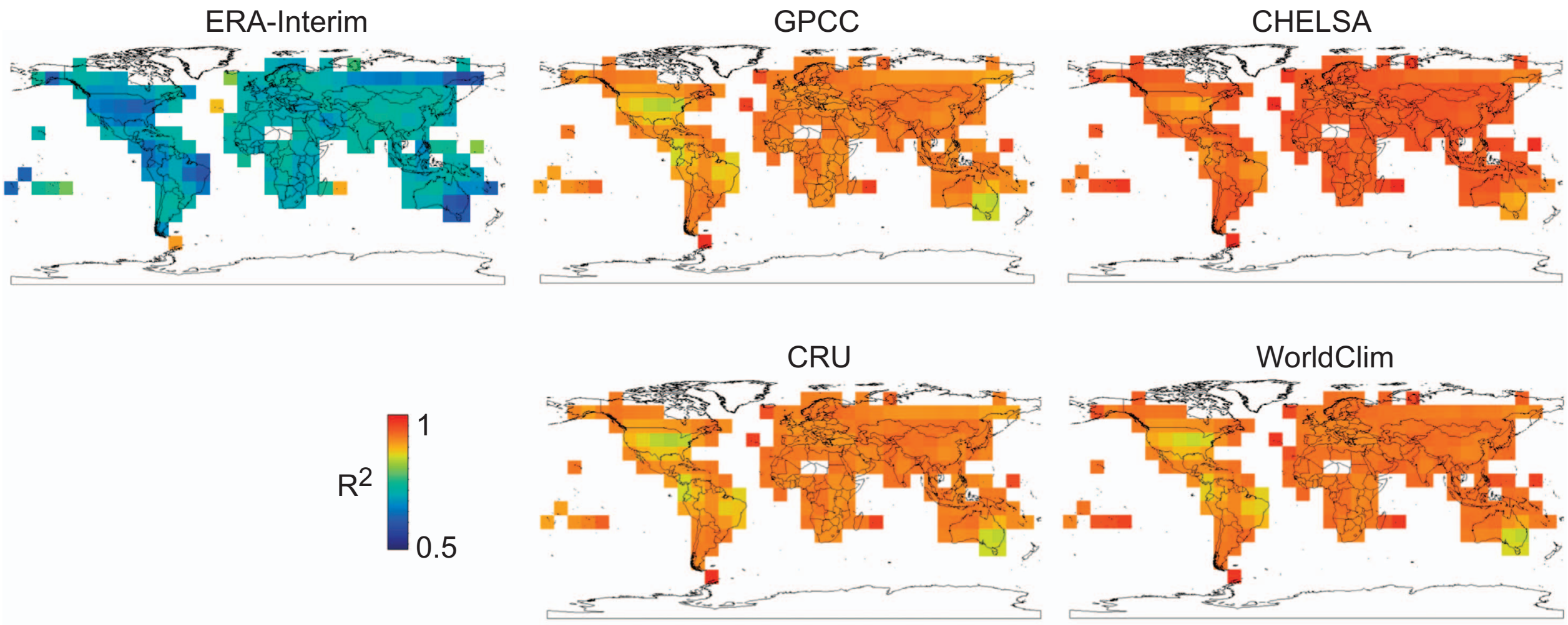

**Figure 2. Small scale comparison of model fit with station data from GHCN[14,15] for annual precipitation sums derived from six different models.** Models have been calculated for each station separately, including a minimum of 16 surrounding stations in a 2° search radius. This illustrates how well modelled precipitation corresponds to measured stations at the scale $<=2°$. The figure shows the mean $R^2$ value for 2° grid cells. The upper row additionally illustrates different steps of the CHELSA algorithm, with ERA-Interim[17] performing worse than GPCC[50], and CHELSA showing the highest fit after including orographic effects.

## Small-scale fit between stations and final climatology

To compare the fit of GHCN stations[14] with the final climatologies, we calculated a linear regression model for each station separately. For each station, the surrounding stations in 2° distance where included (with a minimum of 16 stations). We then regressed the mean annual precipitation measured at the station with the mean annual precipitation derived from six different models that either use GHCN data in their algorithms (CHELSA, CRU[9], WorldClim[8], GPCC[50]) or do not use GHCN directly (ERA-Interim[17]).

From this linear regression approach two inferences can be drawn. First, the comparison between ERA-Interim, GPCC, and CHELSA shows the increase in fit with the specific corrections which are applied in the CHELSA algorithm, as ERA-Interim and GPCC contribute data to CHELSA. Second, the results show how well the stations correspond to the respective models at small spatial scale.

Among the models which use GHCN in their algorithm, CHELSA shows the highest fit between stations and predicted precipitation, with WorldClim, GPCC and CRU showing smaller, but still high fits with the station data (Fig. 2).

## Validation using independent precipitation station data

A statistical comparison with different datasets is complicated by the fact that most gridded temperature and precipitation datasets are parameterized using similar observational data, leading to generally high correlations between climatic reanalyses. To validate the results of the CHELSA algorithm, we identified several independent datasets of various size and temporal extents. None of these have been used in the final bias correction within the algorithm, and we have additionally screened them for duplicates in the GHCN[15], MeteoSwiss, and DWD datasets. As the station data is of different spatial extent, it allows us to validate the accuracy of CHELSA on the global scale, as well as on the very small target scale of 30 arc°. We split the validation in two parts. One part examines the temporal performance of the reanalysis dataset (Table 2) and the other the climatological performance of the dataset (Table 3). For comparison with other reanalysis products, we also calculated a similar validation for the CRU[9] and ERA-Interim[17] datasets. For comparison with other climatologies, we included the CHPclim[52], CRU, ERA-Interim, and WorldClim[8] datasets.

Precipitation validation data:

1. FAO—data: 2,316 stations
2. Mexico—data: 2,950 stations
3. Austria—Ehyd data: 877 stations
4. South Africa—SAEON data: 14 stations
5. Scandinavia—Nordklim data: 11 stations
6. China—CMA data: 241 stations

| | $R^2$ | s.d. | RMSE | s.d. | MAE | s.d. |
|---|---|---|---|---|---|---|
| Austria-CHELSA | 0.43 | 0.15 | 32.49 | 12.71 | 24.03 | 9.47 |
| Austria-CRU | 0.41 | 0.15 | 32.99 | 12.26 | 24.09 | 9.03 |
| Austria-ERA | 0.35 | 0.16 | 34.55 | 13.03 | 25.65 | 9.75 |
| China-CHELSA | 0.61 | 0.17 | 30.85 | 43.87 | 17.61 | 12.65 |
| China-CRU | 0.60 | 0.17 | 36.76 | 35.08 | 21.70 | 26.70 |
| China-ERA | 0.48 | 0.18 | 34.77 | 44.18 | 20.11 | 13.61 |
| FAO-CHELSA | 0.83 | 0.11 | 37.66 | 12.67 | 18.64 | 7.08 |
| FAO-CRU | 0.73 | 0.13 | 48.80 | 18.28 | 22.40 | 8.42 |
| FAO-ERA | 0.51 | 0.11 | 66.81 | 20.76 | 37.82 | 11.52 |
| Mexico-CHELSA | 0.39 | 0.11 | 62.37 | 35.87 | 37.54 | 25.59 |
| Mexico-CRU | 0.35 | 0.11 | 64.21 | 36.92 | 38.49 | 25.86 |
| Mexico-ERA | 0.33 | 0.10 | 66.32 | 39.94 | 41.58 | 28.67 |
| Scandinavia-CHELSA | 0.78 | 0.22 | 24.40 | 14.06 | 16.38 | 9.35 |
| Scandinavia-CRU | 0.79 | 0.22 | 22.79 | 12.07 | 15.07 | 7.80 |
| Scandinavia-ERA | 0.62 | 0.24 | 33.27 | 15.36 | 21.73 | 9.30 |

**Table 2. Reanalysis validation using independent station data.** Values represent means over all months within the validation period for which station data was available.

| | Austria | | China | | FAO | | Mexico | | Scandinavia | |
|---|---|---|---|---|---|---|---|---|---|---|
| | mean | s.d. | mean | s.d. | mean | s.d. | mean | s.d. | mean | s.d. |
| CHELSA-$R^2$ | 0.38 | 0.06 | 0.62 | 0.08 | 0.92 | 0.01 | 0.79 | 0.03 | 0.87 | 0.04 |
| CHPclim-$R^2$ | 0.38 | 0.05 | 0.59 | 0.09 | 0.92 | 0.03 | 0.65 | 0.16 | 0.50 | 0.23 |
| CRU-$R^2$ | 0.34 | 0.02 | 0.64 | 0.10 | 0.48 | 0.26 | 0.45 | 0.11 | 0.86 | 0.11 |
| ERA-$R^2$ | 0.32 | 0.04 | 0.57 | 0.11 | 0.65 | 0.03 | 0.51 | 0.07 | 0.59 | 0.17 |
| WorldClim-$R^2$ | 0.30 | 0.06 | 0.61 | 0.10 | [0.93] | [0.02] | [0.83] | [0.02] | [0.81] | [0.11] |
| CHELSA-RMSE | 25.71 | 1.29 | 17.23 | 2.77 | 24.55 | 3.94 | 28.13 | 17.60 | 14.44 | 5.54 |
| CHPclim-RMSE | 25.70 | 1.28 | 16.86 | 2.38 | 28.97 | 20.23 | 36.23 | 27.98 | 27.55 | 9.97 |
| CRU-RMSE | 26.56 | 0.97 | 15.40 | 2.07 | 58.94 | 26.24 | 43.73 | 24.06 | 13.55 | 3.84 |
| ERA-RMSE | 26.95 | 1.06 | 16.89 | 2.26 | 49.21 | 11.00 | 42.36 | 25.71 | 25.15 | 7.60 |
| WorldClim-RMSE | 27.39 | 1.50 | 16.46 | 2.56 | [20.61] | [3.65] | [24.53] | [14.44] | [16.74] | [6.16] |
| CHELSA-MAE | 19.37 | 0.97 | 12.56 | 2.62 | 13.24 | 1.68 | 16.98 | 12.10 | 9.57 | 3.31 |
| CHPclim-MAE | 19.15 | 0.92 | 12.24 | 2.07 | 14.67 | 15.10 | 22.83 | 20.84 | 19.02 | 6.43 |
| CRU-MAE | 19.78 | 0.83 | 10.09 | 1.89 | 39.96 | 19.72 | 27.56 | 17.32 | 9.56 | 2.84 |
| ERA-MAE | 20.04 | 0.86 | 11.81 | 2.38 | 27.09 | 4.39 | 26.97 | 18.57 | 15.71 | 4.20 |
| WorldClim-MAE | 20.19 | 1.26 | 11.94 | 2.17 | [10.44] | [1.38] | [14.44] | [9.23] | [10.38] | [3.70] |

**Table 3. Climatological validation using independent climate station data.** Values in square brackets indicate that the data has been used in the station interpolation algorithm of the respective dataset and is therefore not valid for an independent evaluation. Values represent means over all months.

## FAO data validation results

The FAO data obtained from the Agromet Group of the Food and Agriculture Organization of the United Nations (FAO) is a collection of 2,316 stations with a good representation in many typically data-sparse regions, but many stations only have a short measuring period. The data is global and duplicates in the GHCN data report have been removed.

CHELSA shows high correspondence of $R^2 = 0.83$ throughout the years with the FAO dataset, while CRU and ERA-interim show considerably lower values ($R^2 = 0.73$ & $R^2 = 0.51$, respectively) (Table 2). The root mean square error (RMSE) and mean absolute error between CHELSA and the FAO data is not significantly different from those of the other validation datasets (with the exception of the SAEON dataset) (Table 2).

For the climatological validation, CHELSA performs similar to CHPclim, and WorldClim (Table 3). All three climatologies, however, already include FAO data in some way, which explains the close fit among the data. CHPclim and WorldClim use them in their station interpolation, and in the case of CHELSA, FAO data are only included in the GPCC data that have been used for the bias correction at the large scale, but not in the GHCN data that have been used in the monthly bias interpolation step. Both CRU and ERA-Interim perform considerably mediocre when compared to FAO data (Table 3). However, this comparison is only partly valid and only shows the increase of fit when station data is included into a precipitation downscaling or interpolation algorithm.

### Mexico data validation results

The Mexico data consist of 2,950 stations with a dense spatial distribution, but with only a short measuring period for many stations.

None of the reanalysis datasets are able to capture the temporal variation in the station dataset well (Table 2). The average $R^2$ of CHELSA only reaches 0.39, which is still slightly better than the performance of CRU and ERA-Interim. The RMSE and MAE values are also lower in their mean, as well as in their standard deviation. The poor performance of all products might be due to the fact that many meteorological stations in this dataset have missing values.

The climatological performance of all models with the Mexico dataset is slightly better than that of the reanalysis dataset (Table 3). WorldClim shows the highest fit with stations, which is not surprising, as WorldClim already includes most of the stations for its original calibration and the Mexico data set is therefore not independent from the WorldClim climatologies. CHELSA shows the second highest fit with the Mexico data and is slightly better than CHPclim in all three metrics ($R^2$, RMSE, MAE). Era-Interim and CRU do not capture the climatological precipitation in this area well, in comparison to the other three datasets.

### Austria Ehyd data validation results

The Ehyd data from the Federal Ministry of Agriculture, Forestry, Environment and Water Management of Austria comprises of a dense net of 877 precipitation stations in Austria.

The overall performance of all precipitation products is low when compared to the Ehyd stations (Table 2). From all models however, CHELSA performs best, with the highest $R^2$, and lowest errors. For the climatologies CHELSA performs second best after CHPclim, but with all climatologies having a comparably low fit with the station data (Table 3). WorldClim shows the lowest fit with station data. In general the overall performance of the climatologies is comparable to that of the reanalysis.

### Skandinavia—Nordklim data

The Nordklim data 1.0 includes observations of twelve climate variables from more than 119 stations in the Nordic region including precipitation and air temperature, in a time span of over 100 years. The data are provided by NORDKLIM/NORDMET on behalf of the National meteorological services in Denmark (DMI), Finland (FMI), Iceland (VI), Norway (DNMI) and Sweden (SMHI). We screened these stations for duplicates in the GHCN dataset and remained with a set of 11 independent stations which we used for the validation.

All reanalysis products track the temporal signal in the data reasonably well, with CRU slightly outperforming CHELSA, and ERA-Interim performing the worst (Table 2). All models, however, show relatively small errors in this region, and are only slightly different in their temporal signal.

CHELSA, WorldClim, and CRU climatologies fit the Nordklim data well, with CHELSA performing better than the other models, despite the fact that Nordklim data are included in WorldClim but not in CHELSA (Table 3). ERA-Interim and CHPclim do not perform well in this region, which is probably due to the larger errors of remote sensing data in arctic regions, on which both models depend.

### China—CMA data

The precipitation data from China comes from the Chinese Meteorological Administration and consists of 241 stations with daily records that are not included in the GHCN dataset. CHELSA and CRU are able to track the temporal signal in precipitation rather well when compared to the CMA data, with ERA-Interim performing less well (Table 2).

For the climatological means, CRU slightly outperforms CHELSA, WorldClim shows a slightly worse performance compared to the former two, and CHPclim and ERA-Interim show the lowest performance in this region, with $R^2$ values slightly below 0.5 (Table 3).

### South Africa—SAEON data validation results

One of the main purposes of CHELSA is the better representation of precipitation gradients at small spatial scales. The SAEON precipitation stations are located in the Jonkershoek valley in South Africa with a strong elevational gradient, from the entrance to the valley to the watershed at the top of the valley. They additionally have a very high interannual variability and have not been included in any global precipitation product. Although the timespan of the dataset is too low to validate the climatological performance, the dataset can be used to track the performance of models in a very complex terrain and a strong seasonality.

For all stations of the SAEON network, CHELSA predicts the temporal variation and the actual precipitation values best compared to the ERA-Interim, CRU and CHIRPS[1] reanalysis products (Fig. 3). All reanalysis products predict the temporal variation in precipitation well, but they differ in the respective errors. All of them underestimate the extremes of precipitation in the region covered by the SAEON data.

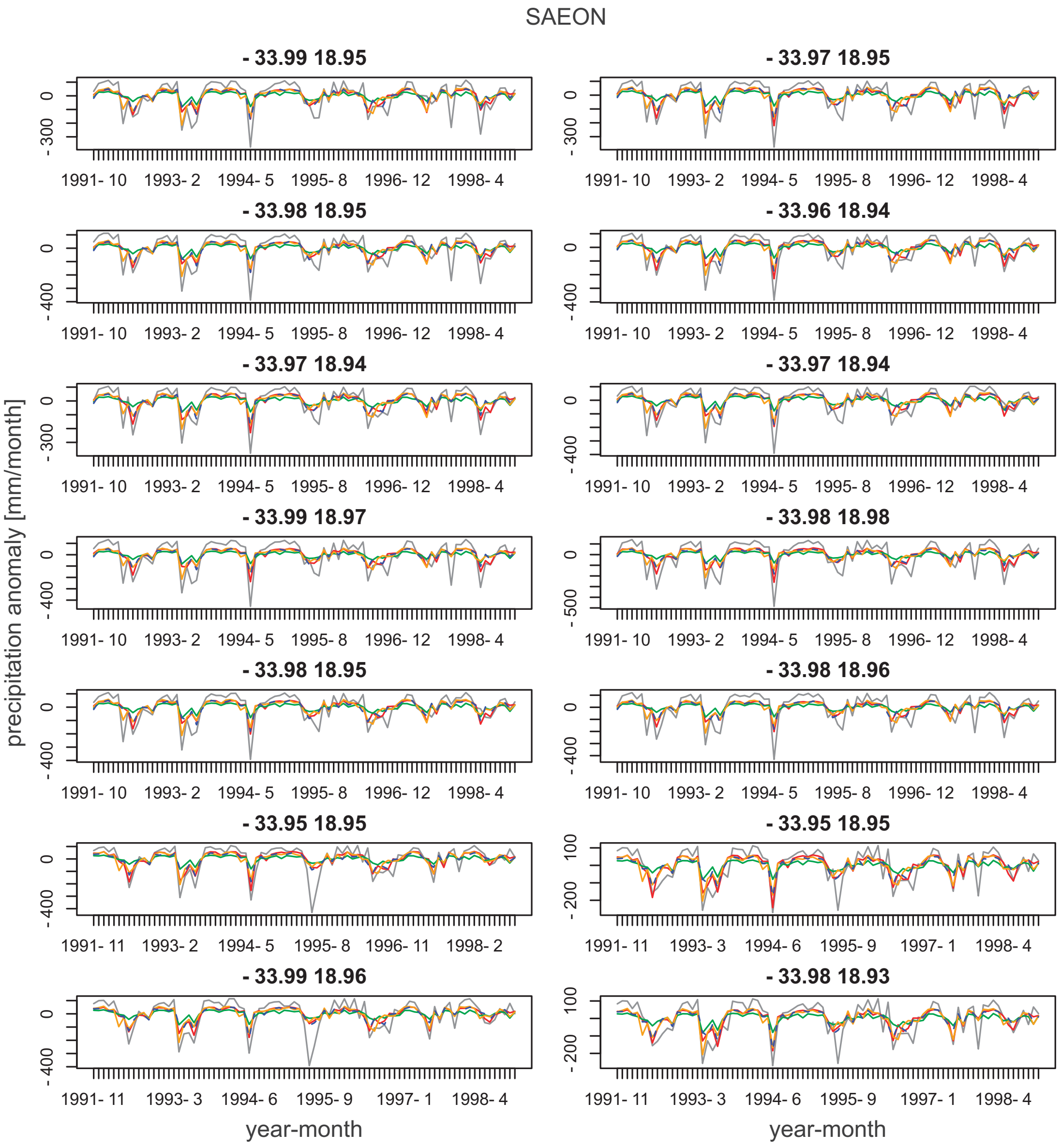

**Figure 3. Temporal precision of the CHELSA reanalysis which forms the basis for the climatologies in a small region of South Africa.** Precipitation anomalies are shown as deviations from the mean precipitation in the respective time period. Grey = station data, red = CHELSA, green = ERA-Interim[17], blue = CRU[9], orange = CHIRPS[1].

## Large-scale spatial comparison of precipitation patterns

To compare our precipitation data with those of other products, we first compared the spatial patterns of precipitation with those of the Tropical Rainfall Measuring Mission (TRMM)[2,3] combined multisatellite product TRMM/TMPA (3B43)[53], CRU[9], WorldClim[8], CHPclim[52], GPCC[50], and ERA-Interim[17]. Figure 4 shows the bias of all mentioned products with the CRU dataset. We used the CRU dataset as a comparison, as it is not included in the other datasets, and therefore the most independent. TRMM/TMPA (3B43), WorldClim, GPCC are all using similar stations, and ERA-Interim is known to exhibit large biases in precipitation. All products, with the exception of ERA-Interim show similar amounts and patterns of biases when compared to CRU data. The bias of CHELSA is lower than that of GPCC and ERA-Interim, the two datasets which have been used in the correction algorithm. The large scale comparison however, only serves as a guideline for the deviation in the above mentioned products in general and cannot be seen as independent validation. For regions in which all models exhibit large biases, we would urge caution in the use of a single precipitation product and would suggest the use of multiple models from various sources.

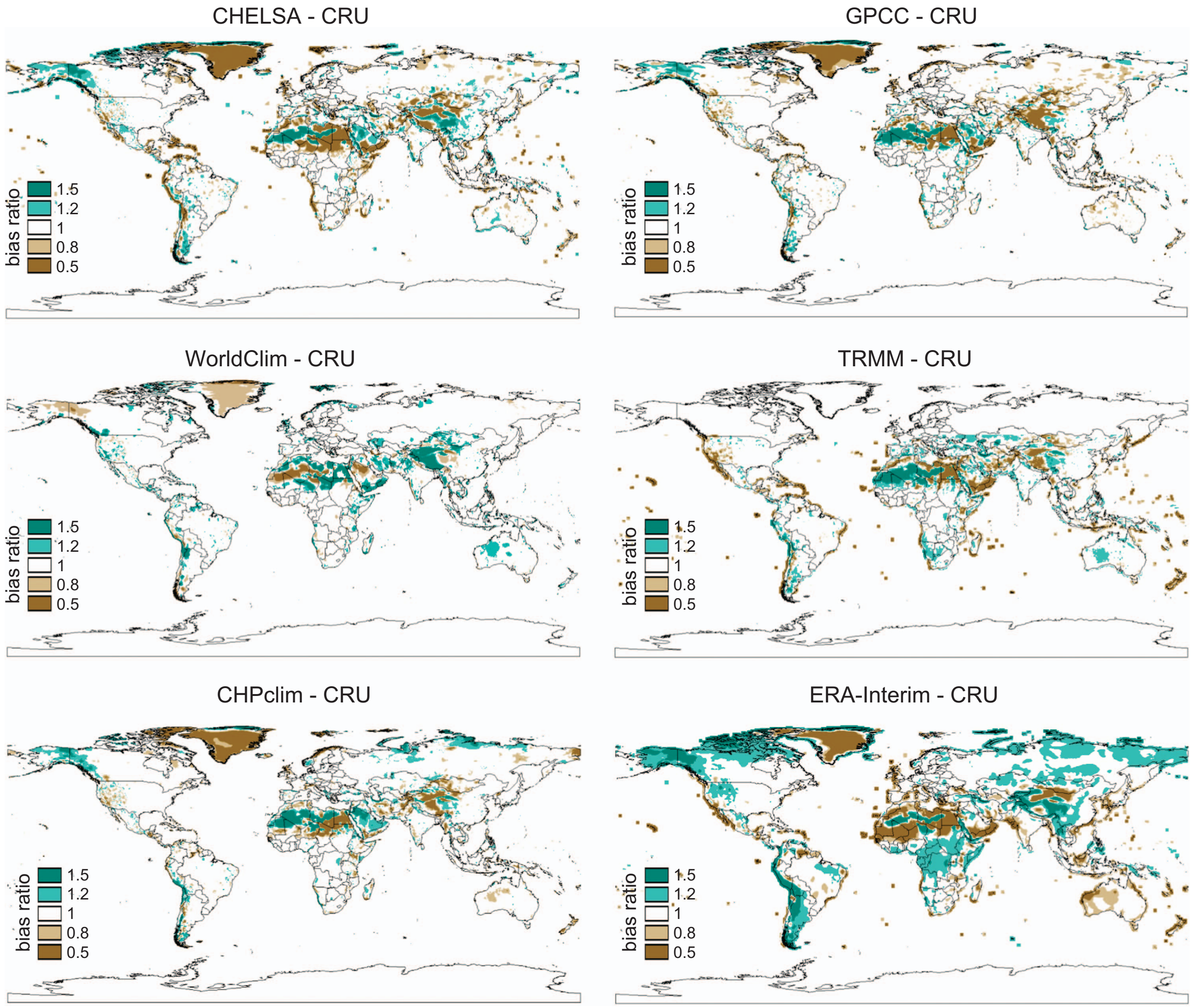

**Figure 4. Bias ratio comparison of annual precipitation sums for six different climatologies with the CRU[9] climatology at the global scale.** All models have a substantial dry bias over Greenland when compared to CRU (TRMM/TMPA (3B43)[53] does not include Greenland). Large differences in bias ratios can be observed in the Atacama, where CHELSA, and GPCC[50] are drier than CRU, and WorldClim[8], CHPclim[52], TRMM/TMPA (3B43), and ERA-Interim[17] are wetter. Also on the Himalaya plateau large differences are visible, with CHELSA, GPCC, and TRMM (3B43) being drier, and WorldClim and ERA-Interim being wetter.

## Small-scale comparison of precipitation patterns

To highlight small scale performance of CHELSA, we compared precipitation patterns of three different models in the topographically and climatically highly complex terrain of Bhutan (Fig. 5). A comparison of the annual precipitation totals between TRMM/TMPA (3B43)[53], WorldClim[8], CHELSA, and the statistical downscaling approach of Böhner[31] shows similar patterns between all models at the mesoscale. The differences at the microscale are, however, severe between CHELSA and Böhner[31] compared to WorldClim[8]. There are only few climate stations in the region of Bhutan, which creates spurious correlations between elevation and precipitation in the ANUSPLIN algorithm of WorldClim. CHELSA and Böhner show a more consistent relation between the terrain features and the resulting precipitation patterns. A comparison with the patterns of cloud formations in this region[54] shows similarities in the patterns where clouds form and where higher precipitation amounts are predicted by CHELSA and Böhner (Fig. 5). Although the formation of clouds does not necessarily coincide with rainfall, there is generally a high correlation between the formation of clouds and the patterns of rainfall especially in topographically complex terrain[55]. We therefore assume that our model is able to capture the topographic heterogeneity of precipitation at the small spatial scale well.

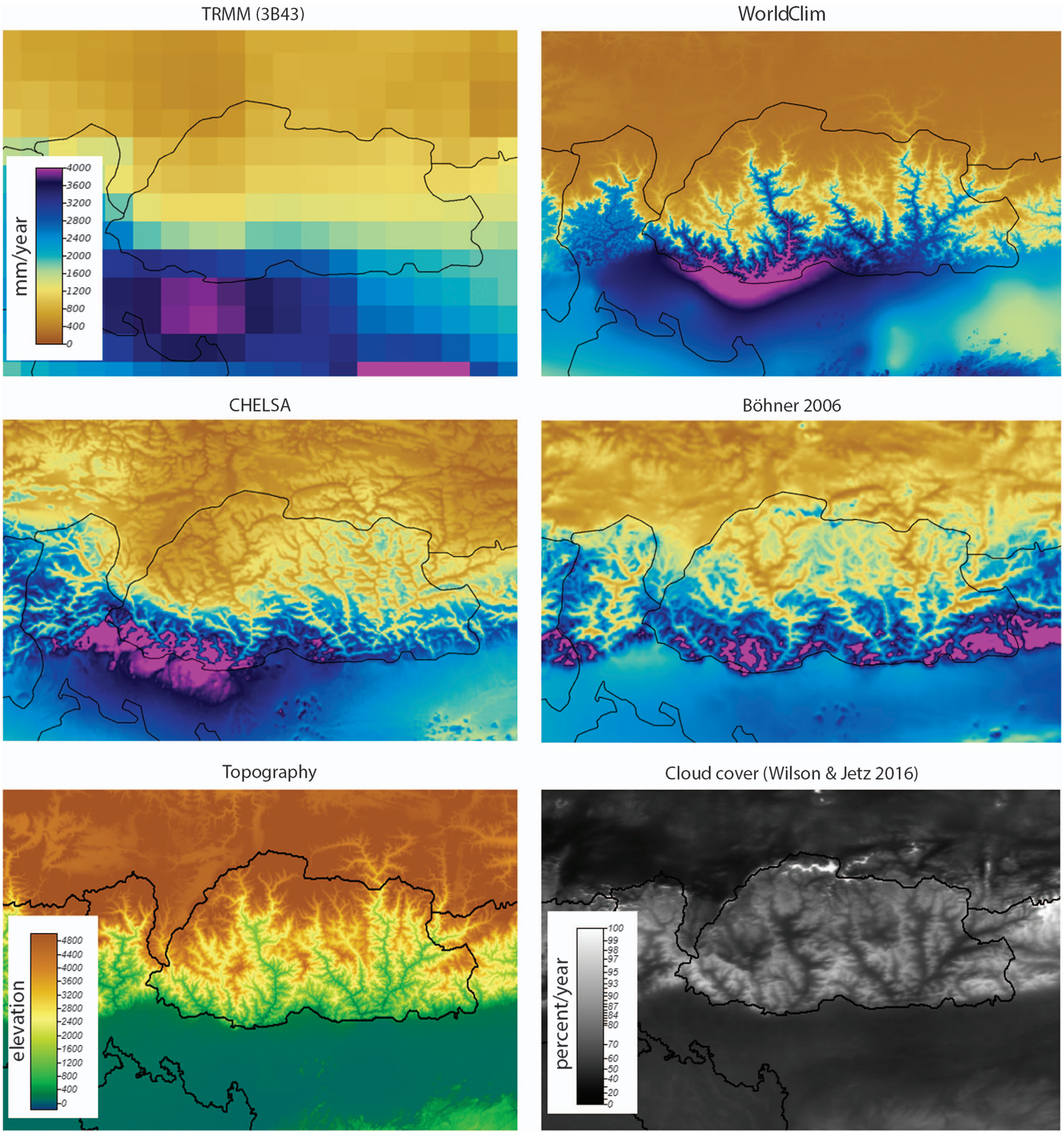

**Figure 5. Comparison of precipitation patterns in the complex terrain of Bhutan (country boundaries in black) between TRMM/TMPA (3B43)[53], WorldClim[8], CHELSA, the statistical downscaling approach of Böhner[31], the topography from GMTED2010[29], and the cloud cover climatology from Wilson & Jetz[54].** In this region, most precipitation falls during the SW-monsoon in the northern summer, when wet air masses from the SW are lifted at the south face of the Himalayas and dry until reaching the Tibetan high plateau. While the mesoscale patterns are in congruence between models, there are clear differences at the microscale. WorldClim[8] predicts wet valleys and dry mountain faces, whereas CHELSA and Böhner[31] predict dry valleys and wet windward exposed mountain faces due to the inclusion of orographic predictors. CHELSA and Böhner[31] are also in closer congruence with the observed distribution of cloud in the area, which shows lower cloud cover in the isolated mountain valleys compared to the wind exposed mountain faces in the south.

| | $R^2$ | s.d. | RMSE | s.d. | MAE | s.d. |
|---|---|---|---|---|---|---|
| CHELSA-GHCN | 0.97 | 0.017 | 1.37 | 0.109 | 1.01 | 0.088 |
| CHELSA-Mexico | 0.81 | 0.036 | 1.95 | 0.103 | 1.43 | 0.100 |
| CHELSA-FAO | 0.96 | 0.037 | 1.50 | 0.296 | 1.02 | 0.209 |
| CHELSA-Nordklim | 0.75 | 0.08 | 1.17 | 0.233 | 0.86 | 0.179 |
| CRUTEM4-GHCN | 0.86 | 0.059 | 3.59 | 0.394 | 2.64 | 0.332 |
| CRUTEM4-Mexico | 0.19 | 0.064 | 2.87 | 0.296 | 2.33 | 0.269 |
| CRUTEM4-FAO | 0.80 | 0.113 | 3.31 | 0.891 | 2.15 | 0.989 |
| CRUTEM4-Nordklim | 0.29 | 0.165 | 2.57 | 0.439 | 1.82 | 0.323 |

**Table 4. Reanalysis validation using independent station data for temperature.** Values represent means over all months within the validation period for which station data were available.

| model | $R^2$ | s.d. | RMSE | s.d. | MAE | s.d. |
|---|---|---|---|---|---|---|
| CHELSA-GHCN | 0.99 | 0.007 | 0.94 | 0.062 | 0.67 | 0.042 |
| CHELSA-Mexiko | 0.89 | 0.023 | 1.47 | 0.075 | 1.08 | 0.080 |
| CHELSA-FAO | 0.98 | 0.014 | 1.11 | 0.094 | 0.78 | 0.079 |
| CHELSA-Nordklim | 0.94 | 0.028 | 0.70 | 0.368 | 0.49 | 0.233 |
| CRUTEM4-GHCN | 0.88 | 0.051 | 3.28 | 0.203 | 2.34 | 0.201 |
| CRUTEM4-Mexiko | 0.20 | 0.060 | 2.84 | 0.283 | 2.30 | 0.249 |
| CRUTEM4-FAO | 0.82 | 0.074 | 3.59 | 0.235 | 2.69 | 0.228 |
| CRUTEM4-Nordklim | 0.32 | 0.178 | 2.51 | 0.399 | 1.82 | 0.316 |

**Table 5. Climatological validation using independent station data.** Values represent means over all months.

## Validation of temperature using independent meteorological stations

We compared CHELSA temperature data to that of MODIS (MOD11C3)[56] and several independent station datasets. Other high resolution products for temperature such as WorldClim do not have the same validation period as CHELSA. A comparison is therefore problematic due to the increase of global temperatures in the last decades[57]. PRISM[11] is geographically restricted to the United States and therefore also not available for global comparisons. As climate station data not directly used by the CHELSA algorithm for temperature, a comparison with station data is possible. We used a set of four different station networks with different temporal and spatial extent for the validation.

### Temperature validation data:

1. FAO—data: 400 stations
2. Mexico—data: 2,915 stations
3. GHCN—data: 6,093 stations
4. Scandinavia—Nordklim data: 32 stations

The downscaled CHELSA data tracks the temperature data well in the GHCN, and FAO datasets, but larger deviations in the Mexico and Nordklim datasets with regard to the $R^2$ values (Table 4). However, the RMSE and MAE of the Nordklim dataset are comparable to those of the two global datasets GHCN and FAO. Only the comparison with the Mexico dataset shows a high RMSE of 1.95, and MAE of 1.43 (Table 4).

The climatologies show a lower RMSE and MAE than the time series when compared to all station data (Table 5). This indicates that although an error exists in the monthly CHELSA reanalysis temperatures, the error does not inflate when these values are averaged. $R^2$ values are also higher for the climatologies, than for the reanalysis values.

## CHELSA—MODIS comparison

Coefficients of determination between MODIS (MOD11C3)[56] and CHELSA temperatures range from 0.95 to 0.99 globally, between GHCN Version 3 and CHELSA temperatures range 0.96 to 0.99 globally, and between MODIS (MOD11C3) and GHCN Version 2 range from 0.83–0.97 (Fig. 6). Both CHELSA and MODIS (MOD11C3) show systematically lower correlations during the northern summer months for the GHCN dataset which might indicate erroneous temperature values in the GHCN dataset. The deviations might, however, also come from the overestimation of temperatures in the arctic by remote sensing data observed as well in the MODIS (MOD11C3) and the ERA-Interim data[17,58]. As MODIS (MOD11C3) and ERA-Interim data are showing a similar bias, we can assume that the deviations either stem from the GHCN dataset or the remote sensing input to ERA-Interim and not the downscaling algorithm we use.

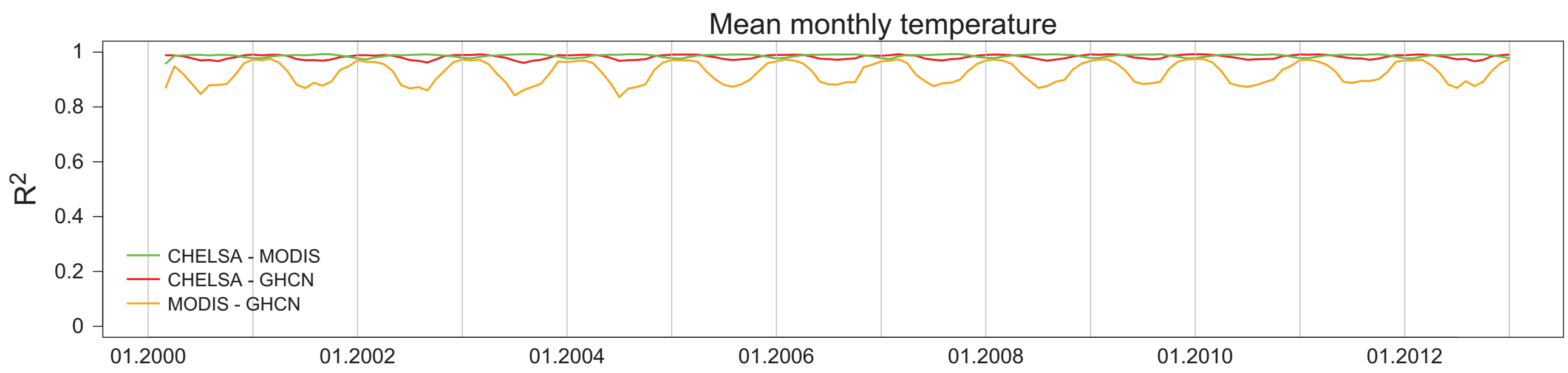

**Figure 6. Temporal comparison of the CHELSA algorithm with GHCN Version 3 (temperature)[14], and MODIS (MOD11C3)[56].** Coefficients of determination give the global correlation between products for a specific month. CHELSA temperatures show significantly higher correlations with GHCN (Wilcoxon Test: W = 23,254, $P < 0.001$). Correlations between CHELSA and MODIS for temperature (mean $R^2 = 0.99$).

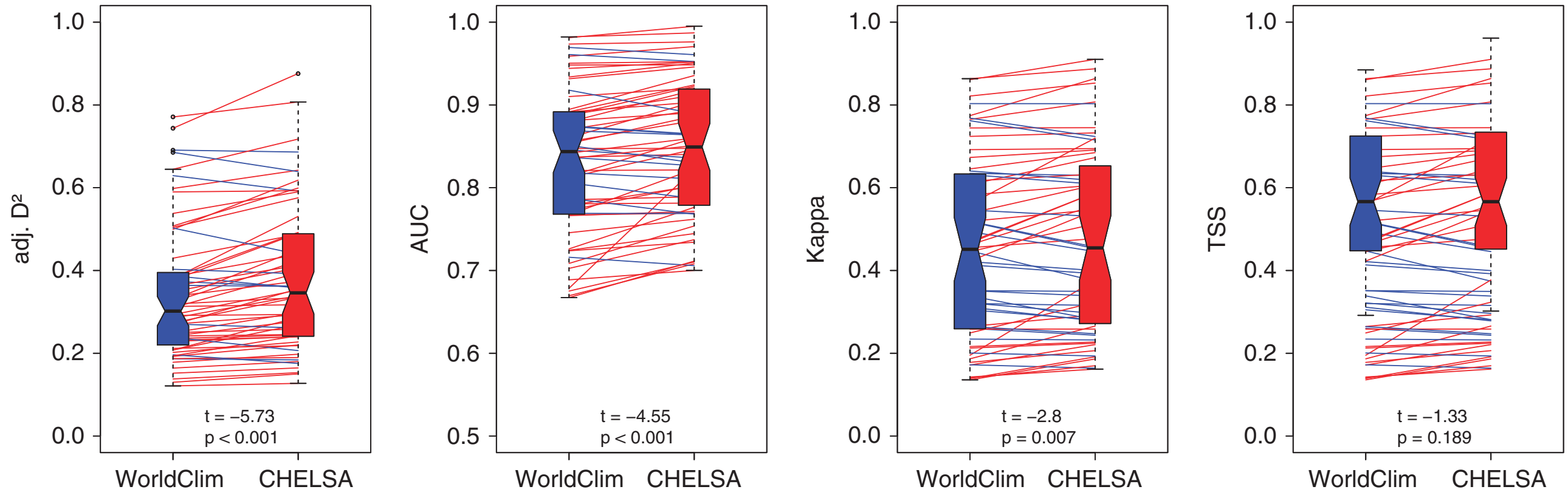

**Figure 7. Comparison of the performance of 67 species distribution models using generalized linear models with WorldClim (blue) and CHELSA (red) precipitation and temperatures in Switzerland as predictors.** Models are compared using paired *t* tests on model performance statistics (adjusted $D^2$, the area under the receiver operation statistics curve (AUC), Kappa statistics, and true skill statistics (TSS)). All performance statistics indicate a higher mean performance of CHELSA over WorldClim[8], with only the TSS statistics not being significant. Red lines indicate models in which SDMs based on CHELSA performed better then SDMs based on WorldClim, blue indicates the opposite relationship. T-values are given for the paired *t*-test comparing WorldClim with CHELSA.

The high spatial correlation between CHELSA and MODIS (MOD11C3)[56] shows that CHELSA is able to predict spatial patterns of temperature distributions well, and additionally accurately predicts the observed values of temperature on a small scale.

## Application example: Performance for species distribution modelling

As we are generally interested in the use of CHELSA climatologies for ecological studies we compare the performance of CHELSA in a species distribution modelling (SDM) approach[59,60] to the most commonly used climate dataset for this purpose: WorldClim[8]. We calculated SDMs for 67 species from Switzerland. We used species from Switzerland as this allows a comparison of performances in areas where climate station density is high in CHELSA and WorldClim, and tests whether the performance improvement of CHELSA is also found in areas with detailed climate data. We modelled species using a generalized linear model with mean annual precipitation and mean annual temperature as predictors. We randomly sampled six times as many pseudo-absence points as presence points and used an inverse weighting approach on the resulting presences and absences. We evaluated the models in a 10-fold cross-validation using the area under the receiver operating characteristic curve (AUC), Kappa statistics[61] and true skill statistic[62] (TSS). AUC and Kappa are traditional test measures between predicted and observed data and usually ranges from 0.5 (AUC) or 0 (Kappa), indicating random fit, to 1, indicating perfect fit. TSS assesses model specificity and sensitivity and ranges from zero (both the specificity and the sensitivity of the model are zero) to 1 (both specificity and sensitivity are 1). Additionally, we calculated the adjusted $D^2$ value which represents the percentage deviance explained by (goodness-of-fit of) the model. Model performance for all 67 species was then compared using a paired *t*-test of all species distribution models.

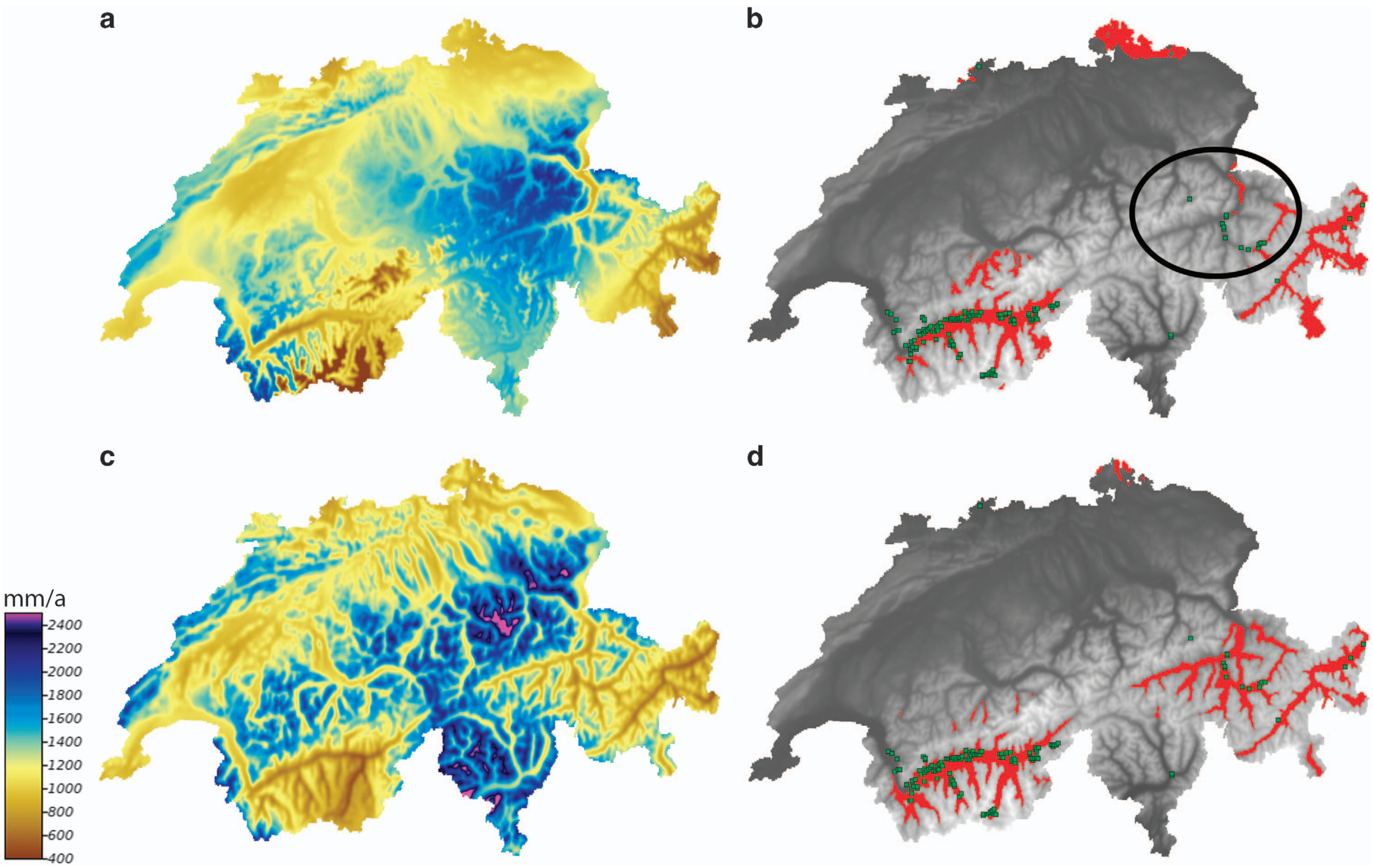

**Figure 8. Comparison of species distribution models based and climate data from WorldClim[8] and CHELSA of *Astragalus monspessulanus* for Switzerland.** Models were calculated using generalized linear models with mean annual precipitation from WorldClim (**a**) and CHELSA (**c**), as well as the mean annual temperature. The distributions based on WorldClim (**b**) and CHELSA (**d**) represent a binary distribution (red) with a threshold of the maximum kappa (WorldClim: threshold = 0.675, AUC = 0.89, $D^2$ = 0.36, Kappa = 0.60, TSS = 0.66, CHELSA: threshold = 0.7155, AUC = 0.91, $D^2$ = 0.47, Kappa = 0.66, TSS = 0.76). Occurrences are marked in green. The circle indicates the area of the dry Rhine valley in eastern Switzerland in which WorldClim overestimates precipitation and therefore does not predict the range of the species correctly in that region (compare with **d**).

The result shows an improvement of the models when using CHELSA over WorldClim data (Fig. 7). All measures show a higher performance of the CHELSA data, although the difference in mean is not significant for the TSS. This shows that even in areas with comparably high station density and good climatic information the CHELSA algorithm improves the spatial prediction of climatic variables and subsequently the modelled distribution of species (Fig. 8).

## Validation results—Conclusions

The validation results in general show that including orographic effects can improve existing climatologies and reanalysis to a degree that the derived analysis (here the SDMs) show increasing accuracies. While CHELSA is an improvement over existing very high-resolution climatologies, it still exhibits errors which we quantified in several ways. The validation of main correction step in the algorithm that includes the orographic wind effects and boundary layer shows that the precipitation at the stations is better captured after the correction than before the downscaling to 30 arc sec resolution. The improvement varies by region and month with the majority of months showing an improvement. Most importantly, the better prediction with regard to SDMs in which precipitation and temperature data are used already indicates that CHELSA might be a substantial improvement over existing products which are currently being employed for such purposes.

## Usage Notes

All CHELSA products are in a geographic coordinate system referenced to the WGS 84 horizontal datum, with the horizontal coordinates expressed in decimal degrees. The CHELSA layer extents (minimum and maximum latitude and longitude) are a result of the coordinate system inherited from the 1-arc-second GMTED2010 data which itself inherited the grid extent from the 1-arc-second SRTM data.

Note that because of the pixel center referencing of the input GMTED2010 data the full extent of each CHELSA grid as defined by the outside edges of the pixels differs from an integer value of latitude or longitude by 0.000138888888 degree (or 1/2 arc-second). Users of products based on the legacy GTOPO30 product should note that the coordinate referencing of CHELSA (and GMTED2010) and GTOPO30 are not the same. In GTOPO30, the integer lines of latitude and longitude fall directly on the edges of a 30-arc-second pixel. Thus, when overlaying CHELSA with products based on GTOPO30 a slight shift of 1/2 arc-second will be observed between the edges of corresponding 30-arc-second pixels.

The dataset is in GEOtiff format. GEOtiff can be viewed using standard GIS software such as:

SAGA GIS—(free) http://www.saga-gis.org/
ArcGIS—https://www.arcgis.com/
QGIS—(free) http://www.qgis.org
DIVA—GIS—(free) http://www.diva-gis.org/
GRASS—GIS—(free) https://grass.osgeo.org/

Grid extent:
Resolution (decimal degrees): 0.0083333333
West extent (minimum X-coordinate, longitude): −180.0001388888
South extent (minimum Y-coordinate, latitude): −90.0001388888
East extent (maximum X-coordinate, longitude): 179.9998611111
North extent (maximum Y-coordinate, latitude): 83.9998611111
Rows: 20,800
Columns: 43,200
The data are feely available under the Creative Commons Licence: CC 0.

## References

1. Funk, C. *et al.* The climate hazards infrared precipitation with stations—a new environmental record for monitoring extremes. *Sci. Data* **2,** 150066 (2015).
2. Biasutti, M., Yuter, S. E., Burleyson, C. D. & Sobel, A. H. Very high resolution rainfall patterns measured by TRMM precipitation radar: seasonal and diurnal cycles. *Clim. Dyn* **39,** 239–258 (2011).
3. Huffman, G. J. *et al.* The TRMM Multisatellite Precipitation Analysis (TMPA): Quasi-Global, Multiyear, Combined-Sensor Precipitation Estimates at Fine Scales. *J. Hydrometeorol.* **8,** 38–55 (2007).
4. Maraun, D. *et al.* Precipitation downscaling under climate change: Recent developments to bridge the gap between dynamical models and the end user. *Rev. Geophys.* **48,** RG3003 (2010).
5. Wood, A. W., Leung, L. R., Sridhar, V. & Lettenmaier, D. P. Hydrologic Implications of Dynamical and Statistical Approaches to Downscaling Climate Model Outputs. *Clim. Change* **62,** 189–216 (2004).
6. Wilby, R. L. *et al.* Statistical downscaling of general circulation model output: A comparison of methods. *Water Resour. Res.* **34,** 2995–3008 (1998).
7. Schmidli, J., Frei, C. & Vidale, P. L. Downscaling from GCM precipitation: a benchmark for dynamical and statistical downscaling methods. *Int. J. Climatol.* **26,** 679–689 (2006).
8. Hijmans, R. J., Cameron, S. E., Parra, J. L., Jones, P. G. & Jarvis, A. Very high resolution interpolated climate surfaces for global land areas. *Int. J. Climatol.* **25,** 1965–1978 (2005).
9. Harris, I., Jones, P. d., Osborn, T. J. & Lister, D. H. Updated high-resolution grids of monthly climatic observations—the CRU TS3.10 Dataset. *Int. J. Climatol.* **34,** 623–642 (2014).
10. Schneider, U. *et al.* GPCC's new land surface precipitation climatology based on quality-controlled in situ data and its role in quantifying the global water cycle. *Theor. Appl. Climatol.* **115,** 15–40 (2013).
11. Daly, C., Taylor, G. H. & Gibson, W. P. The PRISM approach to mapping precipitation and temperature. *in Proc., 10th AMS Conf. on Applied Climatology* 20–23 (1997).
12. Deblauwe, V. *et al.* Remotely sensed temperature and precipitation data improve species distribution modelling in the tropics. *Glob. Ecol. Biogeogr* **25,** 443–454 (2016).
13. Soria-Auza, R. W. *et al.* Impact of the quality of climate models for modelling species occurrences in countries with poor climatic documentation: a case study from Bolivia. *Ecol. Model.* **221,** 1221–1229 (2010).
14. Lawrimore, J. H. *et al.* An overview of the Global Historical Climatology Network monthly mean temperature data set, version 3. *J. Geophys. Res. Atmospheres* **116,** 1–18 (2011).
15. Peterson, T. C. & Vose, R. S. An overview of the Global Historical Climatology Network temperature database. *Bull. Am. Meteorol. Soc* **78,** 2837–2849 (1997).
16. Kalnay, E. *et al.* The NCEP/NCAR 40-Year Reanalysis Project. *Bull. Am. Meteorol. Soc* **77,** 437–471 (1996).
17. Dee, D. P. *et al.* The ERA-Interim reanalysis: configuration and performance of the data assimilation system. *Q. J. R. Meteorol. Soc* **137,** 553–597 (2011).
18. Wilby, R. L. & Wigley, T. M. L. Downscaling general circulation model output: a review of methods and limitations. *Prog. Phys. Geogr.* **21,** 530–548 (1997).
19. Böhner, J., Antonic, O., Böhner, J. & Antonic, O. in *Geomorphometry: Concepts, Software, Applications* (eds Hengl T. & Reuter H. I.) 195–226 (Elsevier Science, 2009).
20. Gerlitz, L., Conrad, O. & Böhner, J. Large-scale atmospheric forcing and topographic modification of precipitation rates over High Asia—a neural-network-based approach. *Earth Syst Dynam* **6,** 61–81 (2015).
21. Berrisford, P. *et al.* The ERA-interim archive. *ERA Rep. Ser* 1–16 (2009).
22. Berrisford, P. *et al.* Atmospheric conservation properties in ERA-Interim. *Q. J. R. Meteorol. Soc* **137,** 1381–1399 (2011).

23. Gao, L. *et al.* Statistical Downscaling of ERA-Interim Forecast Precipitation Data in Complex Terrain Using LASSO Algorithm, Statistical Downscaling of ERA-Interim Forecast Precipitation Data in Complex Terrain Using LASSO Algorithm. *Adv. Meteorol. Adv. Meteorol.* e472741 (2014).
24. Bao, X. & Zhang, F. Evaluation of NCEP-CFSR, NCEP-NCAR, ERA-Interim, and ERA-40 Reanalysis Datasets against Independent Sounding Observations over the Tibetan Plateau. *J. Clim* **26,** 206–214 (2012).
25. Betts, A. K., Köhler, M. & Zhang, Y. Comparison of river basin hydrometeorology in ERA-Interim and ERA-40 reanalyses with observations. *J. Geophys. Res. Atmospheres* **114,** D02101 (2009).
26. Hansen, J., Sato, M. & Ruedy, R. Radiative forcing and climate response. *J. Geophys. Res. Atmospheres* **102,** 6831–6864 (1997).
27. Rolland, C. Spatial and Seasonal Variations of Air Temperature Lapse Rates in Alpine Regions. *J. Clim* **16,** 1032–1046 (2003).
28. Minder, J. R., Mote, P. W. & Lundquist, J. D. Surface temperature lapse rates over complex terrain: Lessons from the Cascade Mountains. *J. Geophys. Res. Atmospheres* **115,** D14122 (2010).
29. Danielson, J. J. & Gesch, D. B. *Global multi-resolution terrain elevation data 2010 (GMTED2010).* (US Geological Survey, 2011).
30. Hunter, R. D. & Meentemeyer, R. K. Climatologically Aided Mapping of Daily Precipitation and Temperature. *J. Appl. Meteorol.* **44,** 1501–1510 (2005).
31. Böhner, J. General climatic controls and topoclimatic variations in Central and High Asia. *Boreas* **35,** 279–295 (2006).
32. Spreen, W. C. A determination of the effect of topography upon precipitation. *Eos Trans. Am. Geophys. Union* **28,** 285–290 (1947).
33. Gao, X., Xu, Y., Zhao, Z., Pal, J. S. & Giorgi, F. On the role of resolution and topography in the simulation of East Asia precipitation. *Theor. Appl. Climatol.* **86,** 173–185 (2006).
34. Basist, A., Bell, G. D. & Meentemeyer, V. Statistical Relationships between Topography and Precipitation Patterns. *J. Clim* **7,** 1305–1315 (1994).
35. Daly, C., Neilson, R. P. & Phillips, D. L. A Statistical-Topographic Model for Mapping Climatological Precipitation over Mountainous Terrain. *J. Appl. Meteorol.* **33,** 140–158 (1994).
36. Sevruk, B. Regional Dependency of Precipitation-Altitude Relationship in the Swiss Alpsin *Climatic Change at High Elevation Sites* (eds Diaz, H. F., Beniston, M. & Bradley, R. S.) 123–137 (Springer Netherlands, 1997).
37. Körner, C. The use of 'altitude' in ecological research. *Trends Ecol. Evol.* **22,** 569–574 (2007).
38. Rotunno, R. & Houze, R. A. Lessons on orographic precipitation from the Mesoscale Alpine Programme. *Q. J. R. Meteorol. Soc* **133,** 811–830 (2007).
39. Weischet, W. & Endlicher, W. *Einführung in die allgemeine Klimatologie* (2008).
40. Roe, G. H. Orographic Precipitation. *Annu. Rev. Earth Planet. Sci.* **33,** 645–671 (2005).
41. Colle, B. A. Sensitivity of Orographic Precipitation to Changing Ambient Conditions and Terrain Geometries: An Idealized Modeling Perspective. *J. Atmospheric Sci* **61,** 588–606 (2004).
42. Sinclair, M. R. A Diagnostic Model for Estimating Orographic Precipitation. *J. Appl. Meteorol.* **33,** 1163–1175 (1994).
43. Smith, R. B. & Barstad, I. A Linear Theory of Orographic Precipitation. *J. Atmospheric Sci* **61,** 1377–1391 (2004).
44. Oke, T. R.. *Boundary layer climates.* Routledge, (2002).
45. Stull, R. B. *An introduction to boundary layer meteorology* 13 (Springer Science & Business Media, 2012).
46. Kållberg, P. *Forecast drift in ERA-Interim* (European Centre for Medium Range Weather Forecasts, 2011).
47. Lafon, T., Dadson, S., Buys, G. & Prudhomme, C. Bias correction of daily precipitation simulated by a regional climate model: a comparison of methods. *Int. J. Climatol.* **33,** 1367–1381 (2013).
48. Arnell, N. W., Hudson, D. A. & Jones, R. G. Climate change scenarios from a regional climate model: Estimating change in runoff in southern Africa. *J. Geophys. Res. Atmospheres* **108,** 4519 (2003).
49. Molteni, F. A. *'historical' approach to the rescaling of ERA-Interim precipitation, internal technical note* (European Centre for Medium Range Weather Forecasts, 2013).
50. Meyer-Christoffer, A. *et al.* GPCC Climatology Version 2015 at 0.25°: Monthly Land-Surface Precipitation Climatology for Every Month and the Total Year from Rain-Gauges built on GTS-based and Historic Data. *Global Precipitation Climatology Centre at Deutscher Wetterdienst* doi: 10.5676/DWD_GPCC/CLIM_M_V2015_025 (2015).
51. Xu, T. & Hutchinson, M. F. New Developments and Applications in the ANUCLIM Spatial Climatic and Bioclimatic Modelling Package. *Env. Model Softw* **40,** 267–279 (2013).
52. Funk, C. *et al.* A global satellite-assisted precipitation climatology. *Earth Syst Sci Data* **7,** 275–287 (2015).
53. Goddard Space Flight Center Distributed Active Archive Center (GSFC DAAC). *TRMM/TMPA 3B43 TRMM and Other Sources Monthly Rainfall Product V7* (2011).
54. Wilson, A. M. & Jetz, W. Remotely Sensed High-Resolution Global Cloud Dynamics for Predicting Ecosystem and Biodiversity Distributions. *PLOS Biol* **14,** e1002415 (2016).
55. Pruppacher, H. R., Klett, J. D. & Wang, P. K. Microphysics of Clouds and Precipitation. *Aerosol Science and Technology* **28,** 381–382 (1998).
56. NASA LP DAAC. *MODIS/Terra Land Surface Temperature and Emissivity Monthly L3 Global 0.05Deg CMG*. NASA EOSDIS Land Processes DAAC, USGS Earth Resources Observation and Science (EROS) Center, (2015).
57. Stocker, T. F. *et al.* IPCC, 2013: climate change 2013: the physical science basis. Contribution of working group I to the fifth assessment report of the intergovernmental panel on climate change (2013).
58. Wan, Z., Zhang, Y., Zhang, Q. & Li, Z.-L. Quality assessment and validation of the MODIS global land surface temperature. *Int. J. Remote Sens.* **25,** 261–274 (2004).
59. Guisan, A. & Zimmermann, N. E. Predictive habitat distribution models in ecology. *Ecol. Model.* **135,** 147–186 (2000).
60. Guisan, A. & Thuiller, W. Predicting species distribution: offering more than simple habitat models. *Ecol. Lett.* **8,** 993–1009 (2005).
61. Warren, D. L., Glor, R. E. & Turelli, M. Environmental Niche Equivalency Versus Conservatism: Quantitative Approaches to Niche Evolution. *Evolution* **62,** 2868–2883 (2008).
62. Allouche, O., Tsoar, A. & Kadmon, R. Assessing the accuracy of species distribution models: prevalence, kappa and the true skill statistic (TSS). *J. Appl. Ecol.* **43,** 1223–1232 (2006).

## Data Citations

1. Karger, D. N. *et al.* *Dryad Digital Repository* http://dx.doi.org/10.5061/dryad.kd1d4 (2017).

## Acknowledgements

M.K. & D.N.K. would like to acknowledge funding from the Swiss National Funds (SNF 147630, SNF 146906). We thank Sergio Maffioletti for implementing the CHELSA algorithm on the science cloud grid computing facility of the University of Zurich. We further thank Stefan Eggenberg and InfoFlora for access to 67 plant species for test modelling.

## Author Contributions

M.K. initiated the project. D.N.K., O.K., and T.K. developed the algorithms in close communication with J.B. R.W.S. compiled the GHCN data and removed the errors. M.K., H.K., P.L., and N.Z. provided the funding for the project. D.N.K. wrote the first draft of the manuscript and all authors contributed significantly to the revisions.

## Additional Information

**Competing interests:** The authors declare no competing financial interests.

**How to cite this article:** Karger, D. N. *et al.* Climatologies at high resolution for the earth's land surface areas. *Sci. Data* 4:170122 doi: 10.1038/sdata.2017.122 (2017).

**Publisher's note:** Springer Nature remains neutral with regard to jurisdictional claims in published maps and institutional affiliations.